# White Holes as the Asymptotic Limit of Evaporating Primordial Black Holes

Jeffrey S. Lee[1]
Gerald B. Cleaver[1,2]

[1]Early Universe Cosmology and Strings Group,
Center for Astrophysics, Space Physics, and Engineering Research
[2]Department of Physics
Baylor University
One Bear Place
Waco, TX 76706

Jeff_Lee@Baylor.edu
Gerald_Cleaver@Baylor.edu

## Abstract

This paper examines the interaction of an intense fermion field with all of the particle species of an attometer primordial black hole's (PBH) high energy Hawking radiation spectrum. By extrapolating to Planck-sized PBHs, it is shown that although Planck-sized PBHs closely simulate the zero absorption requirement of white holes, the absorption probability is not truly zero, and therefore, thermodynamically, Planck-sized primordial black holes are not true white holes.

## 1. Introduction

The Hawking radiation spectrum of a primordial black hole with a Schwarzschild radius in the attometer ($10^{-18}$ m) range is awash with $p$, $\bar{p}$, $e^{\pm}$, $\gamma$, $\nu$, and $\bar{\nu}$. Particles of a surrounding radiation field, incident upon the PBH, will interact with the expelled Hawking radiation and form an accretion cloud of high opacity. Considered in detail in this paper are the interactions between incident fermions and each of the emitted particle species.

Although the Hawking radiation is not self-interactive and does not itself form an accretion cloud, the scattering and particle annihilation that occurs from incident fermions does result in a highly opaque accretion cloud through which particles with energies comparable to the PBH's mass energy cannot retain sufficient energy to have a high absorption probability. It is shown that when extrapolated to Planck-sized PBHs, the absorption probability, although extremely negligible, is non-zero. Consequently, Planck-sized primordial black holes can approximately mimic white holes' zero-absorption characteristic, but they do not achieve it. Thus, in terms of absorptivity, white holes are the asymptotic limit of evaporating primordial black holes.

## 2. Evaporation Times of a Primordial Black Hole

The classical evaporation time $t_{ev}$, of a Schwarzschild black hole with initial Schwarzschild radius $R_{s_o}$ and initial mass $M_o$, obtained from solving $P = -c^2 \frac{dM}{dt}$ (using classical Hawking power) and expressed in MKS units, is:

$$t_{ev} = \frac{5120\pi G_N^2 M_o^3}{\hbar c^4} \tag{1}$$

Since,

$$R_{S_o} = \frac{2G_N M_o}{c^2} \tag{2}$$

we therefore find that,

$$t_{ev} = \frac{640\pi c^2 R_{S_o}^3}{\hbar G_N}, \tag{3}$$

where $G_N$ is the Newtonian Gravitation Constant (6.6738 × $10^{-11}$ $m^3 \cdot kg^{-1} \cdot s^{-2}$). However, significantly shorter evaporation times are calculated when particle production is considered. Crane and Westmoreland [1] have calculated the approximate range of evaporation times to be:

$$\frac{c^4}{75G_N^2 a}\left(R_{s_o}^3 - R_{s_Y}^3\right) \le t_{ev} - t_o \le \frac{c^4}{6G_N^2 a f\left(T_{H_o}\right)}\left(R_{s_o}^3 - R_{s_Y}^3\right), \tag{4}$$

where $R_{s_Y}$ is the Schwarzschild radius of a reference PBH (taken to be 0.6 am) with an evaporation time of $t_o$ (calculated to be one year), $a$ is the radiated power constant (1.06 × $10^{-20}$ $W \cdot m^2$) and $f\left(T_{H_o}\right)$ is a numerical function accounting for particle production [2]. In technicolor and supersymmetric models, $f\left(T_{H_o}\right) \lessapprox 100$.

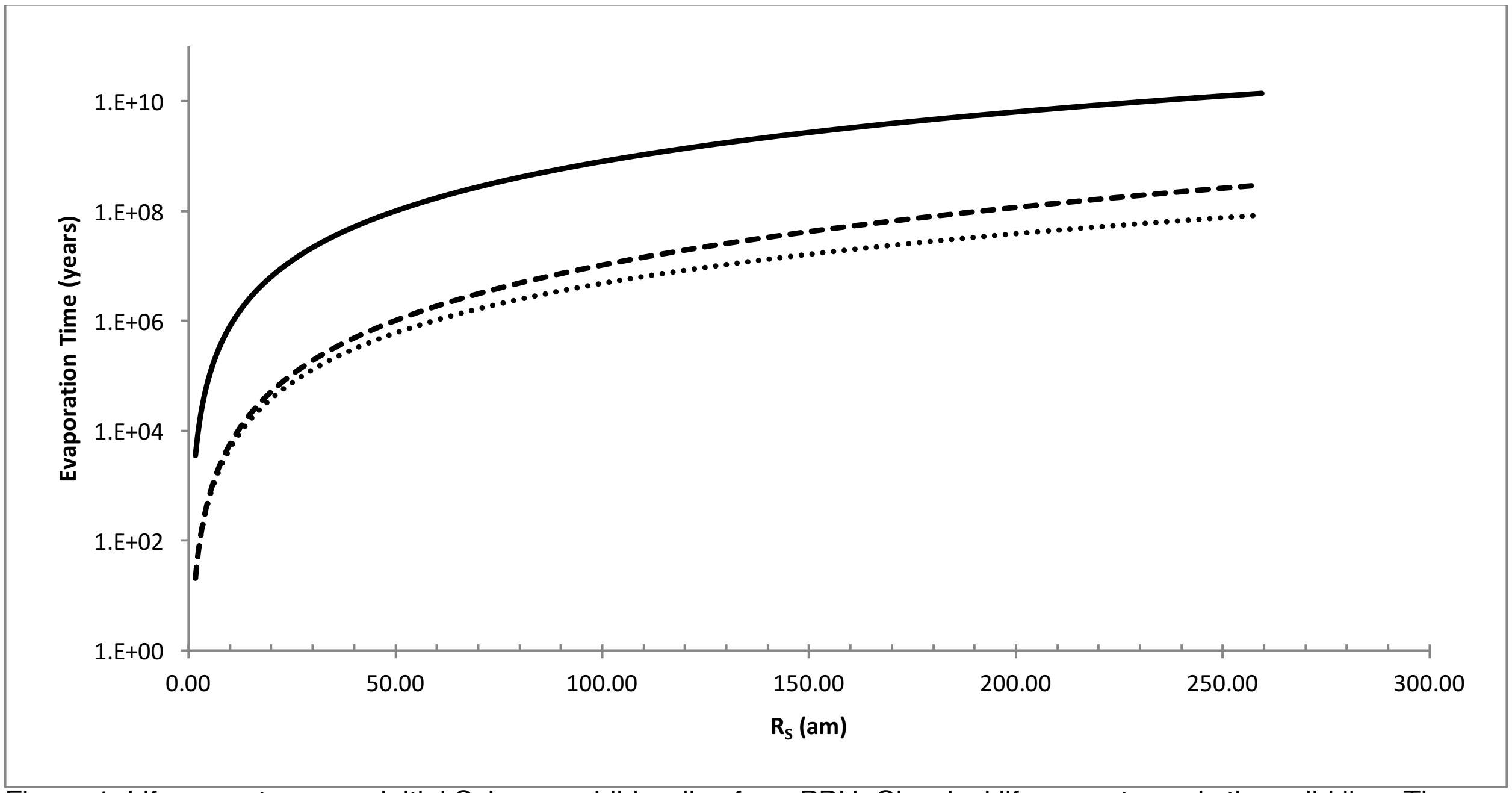

Figure 1: Life expectancy vs. initial Schwarzschild radius for a PBH. Classical life expectancy is the solid line. The lower and upper life expectancy limits are the dotted and dashed lines respectively.

Therefore, the classical evaporation time calculation exaggerates a PBH's life expectancy by 2-3 orders of magnitude.

The above data is shown in Table 1.

Table 1: Schwarzschild radius in attometers, mass in megatons, temperature in gigaelectron-volts, power (petawatts), evaporation rate (kg/s), evaporation time (years), power per unit mass (petawatts per megaton) for PBHs. The results shown here are in accord with Crane and Westmoreland [1].

| $R_S$ (am) | $M$ (MT) | $k_B T$ (GeV) | $P$ (PW) | $P/c^2$ (kg/sec) | $L$ (years) | $P/M$ (PW/MT) |
|---|---|---|---|---|---|---|
| 0.16 | 0.108 | 98.1 | 5519 | 61.4 | 0.04 | 51101.85185 |
| 0.3 | 0.202 | 52.3 | 1527 | 17 | 0.12 | 7559.405941 |
| 1 | 0.673 | 15.7 | 129 | 1.43 | 5 | 191.679049 |
| 1.5 | 1.01 | 10.5 | 56.2 | 0.626 | 16-17 | 55.64356436 |
| 2 | 1.35 | 7.85 | 31.3 | 0.348 | 39-40 | 23.18518519 |
| 2.5 | 1.68 | 6.28 | 19.8 | 0.221 | 75-80 | 11.78571429 |
| 3 | 2.02 | 5.23 | 13.7 | 0.152 | 131-140 | 6.782178218 |
| 6 | 4.04 | 2.62 | 3.26 | 0.0362 | 1042-1177 | 0.806930693 |
| 7 | 4.71 | 2.24 | 2.36 | 0.0262 | 1661-1903 | 0.501061571 |
| 10 | 6.73 | 1.57 | 1.11 | 0.0123 | 4843-5783 | 0.164933135 |

MacGibbon has calculated the fractions of total flux, total power, and kinetic energy transported by each particle species.

Table 2: Fractions of total power transported by the emitted species, including statistical errors [2].

| T (GeV) | | $p\bar{p}$ | $e^{\pm}$ | $\gamma$ | $\nu\bar{\nu}$ |
|---|---|---|---|---|---|
| 0.3 | $P_{TOT} = 2.17 \pm 0.05 \times 10^{23}$ s$^{-1}$ | 10.23% | 20.65% | 23.09% | 46.03% |
| | (% of $P_{TOT}$) | (±0.31%) | (±0.47%) | (±0.51%) | (±0.99%) |
| | Jet Products | 10.23% | 11.18% | 21.37% | 30.35% |
| | (% of $P_{TOT}$) | (±0.31%) | (±0.94%) | (±0.64%) | (±1.34%) |
| 1 | $P_{TOT} = 3.03 \pm 0.05 \times 10^{24}$ s$^{-1}$ | 8.79% | 20.19% | 24.14% | 46.89% |
| | (% of $P_{TOT}$) | (±0.19%) | (±0.30%) | (±0.35%) | (±0.66%) |
| | Jet Products | 8.79% | 12.14% | 22.20% | 32.15% |
| | (% of $P_{TOT}$) | (±0.19%) | (±0.46%) | (±0.42%) | (±0.69%) |
| 10 | $P_{TOT} = 3.56 \pm 0.09 \times 10^{26}$ s$^{-1}$ | 11.22% | 18.71% | 24.73% | 45.33% |
| | (% of $P_{TOT}$) | (±0.20%) | (±0.23%) | (±0.25%) | (±0.43%) |
| | Jet Products | 11.22% | 12.13% | 23.12% | 32.86% |
| | (% of $P_{TOT}$) | (±0.20%) | (±0.59%) | (±0.34%) | (±0.71%) |
| 50 | $P_{TOT} = 9.79 \pm 0.28 \times 10^{27}$ s$^{-1}$ | 11.36% | 18.79% | 24.77% | 45.08% |
| | (% of $P_{TOT}$) | (±0.21%) | (±0.30%) | (±0.41%) | (±0.92%) |
| | Jet Products | 11.36% | 12.37% | 23.26% | 33.75% |
| | (% of $P_{TOT}$) | (±0.21%) | (±0.53%) | (±0.50%) | (±1.22%) |
| 100 | $P_{TOT} = 3.91 \pm 0.12 \times 10^{28}$ s$^{-1}$ | 11.10% | 18.93% | 24.70% | 45.27% |
| | (% of $P_{TOT}$) | (±0.17%) | (±0.19%) | (±0.25%) | (±0.52%) |
| | Jet Products | 11.10% | 12.37% | 23.22% | 24.52% |
| | (% of $P_{TOT}$) | (±0.17%) | (±0.44%) | (±0.36%) | (±0.84%) |

Table 3: Fractions of total flux transported by the emitted species, including statistical errors [2].

| T (GeV) | | $p\bar{p}$ | $e^{\pm}$ | $\gamma$ | $\nu\bar{\nu}$ |
|---|---|---|---|---|---|
| 0.3 | $\dot{N}_{TOT}$ = 1.13 ± 0.02 × $10^{24}$ GeV$^{-1}$·s$^{-1}$ (% of $\dot{N}_{TOT}$) | 1.75% (±0.04%) | 20.62% (±0.23%) | 20.33% (±0.23%) | 57.30% (±0.49%) |
| | Jet Products (% of $\dot{N}_{TOT}$) | 1.75% (±0.04%) | 18.24% (±0.27%) | 19.89% (±0.25%) | 52.69% (±0.55%) |
| 1 | $\dot{N}_{TOT}$ = 1.05 ± 0.01 × $10^{25}$ GeV$^{-1}$·s$^{-1}$ (% of $\dot{N}_{TOT}$) | 1.75% (±0.03%) | 20.15% (±0.14%) | 20.89% (±0.15%) | 58.20% (±0.31%) |
| | Jet Products (% of $\dot{N}_{TOT}$) | 1.75% (±0.03%) | 18.89% (±0.16%) | 20.35% (±0.16%) | 54.13% (±0.37%) |
| 10 | $\dot{N}_{TOT}$ = 3.89 ± 0.08 × $10^{26}$ GeV$^{-1}$·s$^{-1}$ (% of $\dot{N}_{TOT}$) | 2.18% (±0.03%) | 19.62% (±0.13%) | 22.19% (±0.14%) | 56.01% (±0.28%) |
| | Jet Products (% of $\dot{N}_{TOT}$) | 2.18% (±0.03%) | 19.25% (±0.13%) | 22.00% (±0.20%) | 44.06% (±0.29%) |
| 50 | $\dot{N}_{TOT}$ = 4.28 ± 0.09 × $10^{27}$ GeV$^{-1}$·s$^{-1}$ (% of $\dot{N}_{TOT}$) | 2.30% (±0.02%) | 19.64% (±0.09%) | 22.09% (±0.09%) | 55.97% (±0.19%) |
| | Jet Products (% of $\dot{N}_{TOT}$) | 2.30% (±0.02%) | 19.49% (±0.09%) | 22.02% (±0.10%) | 55.59% (±0.19%) |
| 100 | $\dot{N}_{TOT}$ = 1.12 ± 0.03 × $10^{28}$ GeV$^{-1}$·s$^{-1}$ (% of $\dot{N}_{TOT}$) | 2.37% (±0.02%) | 19.63% (±0.09%) | 23.13% (±0.10%) | 55.88% (±0.21%) |
| | Jet Products (% of $\dot{N}_{TOT}$) | 2.37% (±0.02%) | 19.50% (±0.44%) | 22.07% (±0.11%) | 55.59% (±0.21%) |

Table 4: Average kinetic energies in GeV of the emitted species, including statistical errors [2].

| T (GeV) | $p\bar{p}$ | $e^{\pm}$ | $\gamma$ | $\nu\bar{\nu}$ |
|---|---|---|---|---|
| 0.3 | 0.190 (±0.001) | 0.192 (±0.001) | 0.219 (±0.001) | 0.155 (±0.001) |
| 1 | 0.515 (±0.001) | 0.289 (±0.001) | 0.335 (±0.001) | 0.238 (±0.001) |
| 10 | 3.781 (±0.001) | 0.872 (±0.002) | 1.021 (±0.001) | 0.741 (±0.001) |
| 50 | 10.340 (±0.023) | 2.187 (±0.009) | 2.565 (±0.013) | 1.843 (±0.018) |
| 100 | 15.450 (±0.040) | 3.367 (±0.006) | 3.899 (±0.005) | 2.829 (±0.012) |

## 3. PBH Absorption of Incident Radiation

Neglecting any interaction with Hawking radiation, the de Broglie wavelength of an incoming particle needs to be comparable to the Schwarzschild radius, in order to have a significant probability of being absorbed by a PBH. For a wavelength of, for instance, 1 am, this corresponds to an energy of 1.24 TeV.

The non-relativistic Planckian spectrum is given by:

$$B_\lambda(T) = \frac{2hc^2}{\lambda^5}\left(\frac{1}{\exp\left(\frac{hc}{\lambda k_B T}\right) - 1}\right) \tag{5}$$

$k_B$ is the Boltzmann Constant. For the G-type star and a 1 am PBH, $\log\left(B_{10^{-18}\,\mathrm{m}}(T \approx 5778\mathrm{K})\right) \approx -10^{12}$. Of the approximately $10^{62}$ photons emitted to date by the sun, none have had sufficiently short wavelengths to be absorbed by a primordial black hole. From Wien's Law, the temperature required to emit radiation dominated by $10^{-18}$ m photons is approximately $10^{15}$ K; the CMB was at this temperature during the Quark Epoch of the Radiation Era ($10^{-10}$ s after the Big Bang).

Since the electric charge of an attometer-sized PBH would be radiated away in a time that is much shorter than the evaporation time [3], the Schwarzschild metric is apropos. Additionally, if fermions are considered as the particles incident upon the horizon of a PBH, the relativistic particle shower in which the PBH would be engulfed, would constitute a high intensity Dirac field.

### 3.1 The Dirac Field in Schwarzschild Spacetime

For convenience and clarity, $\hbar = c = k_B = 1$. The Schwarzschild metric, in spherical coordinates, is given by:

$$ds^2 = \left(1-\frac{1}{r}\right)dt^2 - \frac{dr^2}{\left(1-\frac{1}{r}\right)} - r^2 d\theta^2 - r^2 \sin^2\theta d\phi^2 \tag{6}$$

The Schwarzschild horizon is located at $r = 1$ (the location of the singularity). For incident particles, the relevant region for absorption is $r \to 1$.

In curved spacetime, the Dirac equation is:

$$\left(i\gamma^a e_a^\mu \left(D_\mu + \Gamma_\mu\right) + m\right)\psi = 0, \tag{7}$$

where $D_\mu = \partial_\mu - \frac{i}{4}\omega_\mu^{ab}\sigma_{ab}$ is the covariant derivative for fermion fields, and $e_a^\mu$ is the vierbein given by (8), $\sigma_{ab} = \frac{i}{2}\left[\gamma_a, \gamma_b\right]$ is the commutator of the Dirac matrices, and $\omega_\mu^{ab}$ are the spin correction components.

$$e_a^\mu = \begin{pmatrix} f^{-\frac{1}{2}} & 0 & 0 & 0 \\ 0 & f^{\frac{1}{2}}\sin\theta\cos\phi & r^{-1}\cos\theta\cos\phi & -r^{-1}\csc\theta\sin\phi \\ 0 & f^{\frac{1}{2}}\sin\theta\sin\phi & r^{-1}\cos\theta\sin\phi & -r^{-1}\csc\theta\cos\phi \\ 0 & f^{\frac{1}{2}}\cos\theta & -r^{-1}\sin\theta & 0 \end{pmatrix} \tag{8}$$

$\gamma_a$ are the Dirac matrices. The spin is specified by,

$$\Gamma_\mu = \frac{1}{2}\left[\gamma^a, \gamma^b\right] e_a^\nu e_{b\nu;\mu} \tag{9}$$

$e_{b\nu;\mu} = \partial_\mu e_{b\nu} - \Gamma_{\nu\mu}^\kappa e_{b\kappa}$ is the covariant derivative of $e_{b\nu}$ and $\Gamma_{\nu\mu}^\kappa$ are the Christoffel symbols. It can be shown that the Dirac radial equation for a general spacetime metric is:

$$\frac{d^2G}{dx^2} + \left[\frac{d}{dx}\left(\frac{f^{\frac{1}{2}}}{1+\lambda f^{\frac{1}{2}}}\frac{k}{r}\right) - \frac{f}{\left(1+\lambda f^{\frac{1}{2}}\right)^2}\left(\frac{k}{r}\right)^2 + \varepsilon^2\left(\frac{1-\lambda f^{\frac{1}{2}}}{1+\lambda f^{\frac{1}{2}}}\right)\right]G = 0 \tag{10}$$

From $f = \rho^2$:

$$\frac{d^2G}{dx^2} + \left[\frac{d}{dx}\left(\frac{\rho}{1+\lambda\rho}\frac{k}{r}\right) - \frac{\rho^2}{(1+\lambda\rho)^2}\left(\frac{k}{r}\right)^2 + \varepsilon^2\left(\frac{1-\lambda\rho}{1+\lambda\rho}\right)\right]G = 0. \tag{11}$$

Using the convenient substitution $dx = \frac{1+\lambda\rho}{\rho^2}dr$ and the relationship $\lambda = \frac{\mu}{\varepsilon}$, the inverse of the vierbein $e_a^\mu$ is [4]:

$$\begin{aligned}&\frac{d^2G}{dx^2} + \frac{1}{\rho^2}\frac{d\rho^2}{dr}\frac{dG}{dr} + (1+\lambda\rho)\frac{d}{dr}\left((1+\lambda\rho)^{-1}\right)\frac{dG}{dr} + \\ &\left[\frac{1}{\rho^2}\frac{k}{r}\frac{d\rho}{dr} + \left(\frac{1+\lambda\rho}{\rho}\frac{k}{r}\right)\frac{d}{dr}\left(\frac{1+\lambda\rho}{\rho}\right) + \frac{1}{\rho}\frac{d}{dr}\left(\frac{k}{r}\right) - \left(\frac{k}{r}\right)^2\left(\frac{1}{\rho}\right) + \frac{\varepsilon^2}{\rho^4} - \left(\frac{\mu}{\rho}\right)^2\right]G = 0\end{aligned} \tag{12}$$

$\Im(G)$, obtained through the WKB approximation, contains wave functions for both the incident and reflected waves.

For $r \sim 1$ (i.e., in the purlieu of the horizon), the interference between incident (the first term in eq. (13)) and reflected (the second term in eq. (13)) waves is:

$$\Im(G(r)) \approx \exp(-i\varepsilon\ln(r-1)) + |R|\exp(i\varepsilon\ln(r-1)). \tag{13}$$

After completion of a 2π rotation in the complex *z*-plane, $\Im(G)$ attains the value $\Im(G(r,2\pi)) \neq \Im(G)$ on its Riemannian surface.

$$G(r,2\pi) \sim \varsigma\exp(-i\varepsilon\ln(r-1)) + \frac{|R|}{\varsigma}\exp(i\varepsilon\ln(r-1)). \tag{14}$$

$|R|$ is the reflection coefficient. $\varsigma = \exp(-2\pi\varepsilon) < 1$. Thus, $|R| = \varsigma = \exp(-2\pi\varepsilon)$. Since the wave will either be reflected or absorbed, therefore, $P_{\text{ref}} + P_{\text{abs}} = 1$ and $P_{\text{ref}} = |R|^2$, and thus, $P_{\text{abs}} = 1 - \exp(-4\pi\varepsilon)$.

Greater detail can be found by consulting [4].

In terms of the Hawking temperature $T_H$, the absorption probability becomes:

$$P_{\text{abs}} = 1 - \exp\left(-\frac{\varepsilon}{T_H}\right), \tag{15}$$

where $\varepsilon$ is the energy of the re-inflating particles.

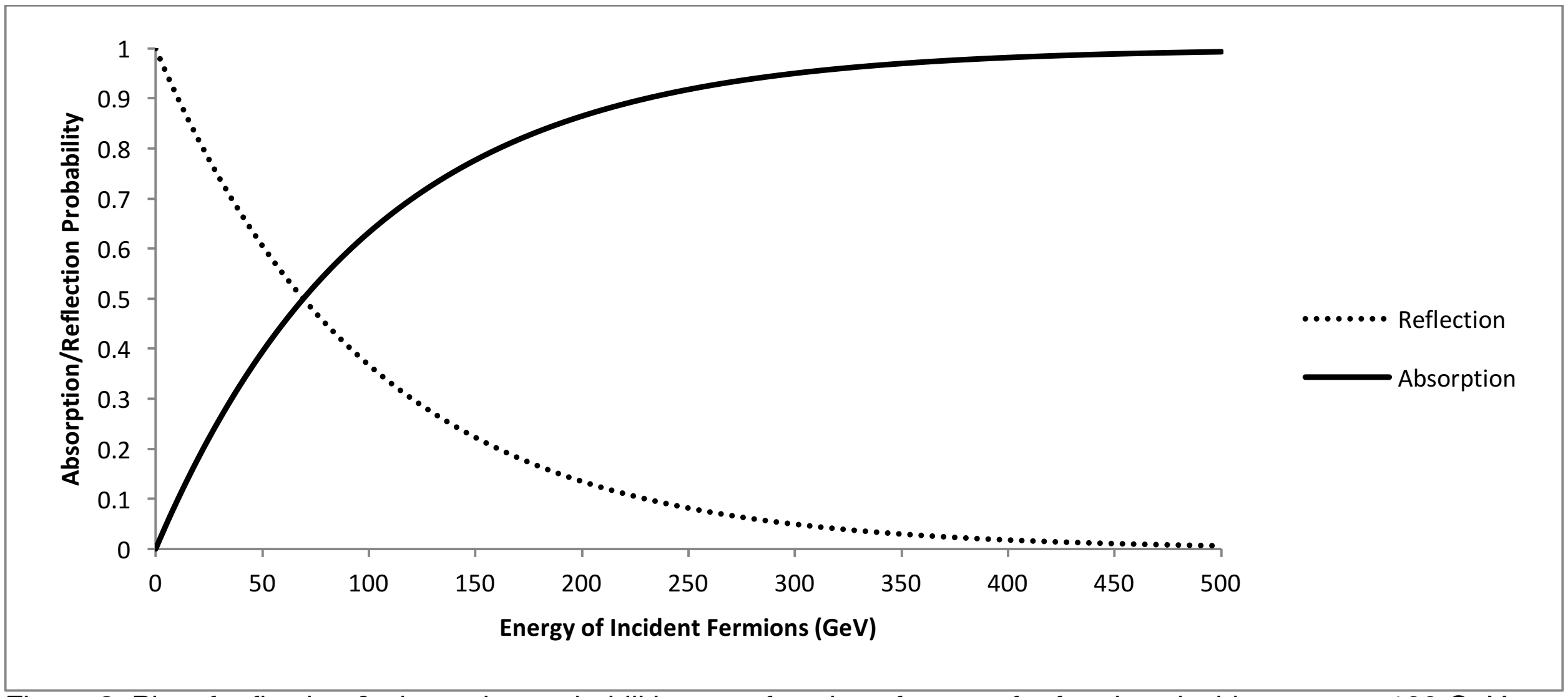

Figure 2: Plot of reflection & absorption probabilities as a function of energy for fermions incident upon a 100 GeV PBH. The reflection and absorption probabilities are equal when $\varepsilon = T_H \ln 2$ (about 70 GeV for a 100 GeV PBH).

From Figure 2, it is desirable that incident fermions arbitrarily have an energy of at least $5k_B T_H$ , and therefore, an absorption probability of 0.9933.

### 3.2 The Effect on Absorption of Uncertainty in Position

If an incident fermion field of nearly infinite pseudorapidity is considered, the Heisenberg Uncertainty Principle, $Dx \cdot Dp \geq \frac{\hbar}{2}$ (where $\hbar$ is the reduced Planck's constant), significantly increases the required energy of the incident fermions for a given absorption probability. If the incoming fermions have an uncertainty in position of, for instance, $0.1R_S$, the corresponding uncertainty in energy is $DE = \frac{5\hbar c}{R_S} = 20\pi k_B T_H$ . For a 1 attometer PBH, this corresponds to $\sim 1\,\mathrm{TeV}$ .

In this scenario, in order to have a low reflection probability (< 0.1), incident fermions would need to have a minimum of $(20\pi + 5)k_B T_H \sim 67.8 k_B T_H$ . For a reflection probability < 0.01, the required energy is $6333 k_B T_H$ . For a 100 GeV PBH, the incident fermion field would need an average energy of $\sim 63\,\mathrm{TeV}$ , which is two orders of magnitude greater than the rest energy of the PBH[i].

### 3.3 Incident Fermion Flux: Non-interaction

Clearly, if a PBH were to increase in mass, the rate of energy absorption from incoming particles must exceed the energy loss rate by Hawking radiation.

---

[i] About four times greater than the upper limit of the current energy capability of the Large Hadron Collider (LHC).

In the simplest approximation, the interaction between the incident radiation field and the Hawking radiation was neglected. However, this simplistic approach significantly underestimates the fermion energy requirement for absorption due to energy losses sustained from scattering and particle annihilation.

### 3.4 Incident Fermion Flux: Interaction

The non-interaction model for determining the minimum necessary flux for a PBH to retain its mass does not account for particle-particle interaction between the outgoing Hawking radiation and the incoming radiation field.

The most elementary approach to particle interaction is to treat the entire Hawking radiation spectrum as photonic. However, the instantaneous Hawking radiation is composed of $p\bar{p}$, $e^{\pm}$, $\gamma$, and $\nu\bar{\nu}$, each particle species contributing substantially to the emergent flux and total power.

A black hole with angular velocity *Ω*, electric potential *V*, and surface gravity $\kappa$, which emits particles of spin *s*, charge *q*, axial quantum number $n\hbar$, absorption probability $\Gamma_s$, has a particle emission rate for particles with energies between *E* and *E + dE*, per degree of particle freedom given by:

$$\dot{N} = \frac{\Gamma_s dE}{2\pi\hbar}\left[\exp\left(\frac{E - n\hbar\Omega - qV}{\hbar\kappa/2\pi c}\right) - (-1)^{2s}\right]^{-1}. \tag{16}$$

Since for a PBH, the electric potential goes to zero very rapidly, and most of its lifetime is spent with $\Omega = 0$, eq. (16) strongly resembles the thermal emission of a blackbody.

Photons, massless neutrinos and very low mass neutrinos contribute very high Hawking fluxes at all PBH temperatures. Often, these particles are considered to tunnel quantum mechanically through the event horizon [5]. More specifically, virtual particle pairs are being spontaneously produced, by the gravitational field, in the region of the event horizon [6]. Particle annihilation is prevented if the virtual pair's wavelength (i.e., the separation of the particles) is approximately the Schwarzschild radius. The observed thermal radiation of a black hole is actually a positive-energy particle escaping to infinity after a classically prohibited negative-energy particle quantum mechanically tunnels through the event horizon to the black hole's interior.

The exponential dependency of $\dot{N}$ on the energy assures some contribution of massive species at all energies. A black hole will conserve all of the associated quantum numbers because it acts as a source of any massless gauge group [7], including the SU(3) color gauge field; thus, the emission of both Strongly Interacting Massive Particles (SIMPs) and Weakly Interacting Massive Particles (WIMPs) is expected for temperatures beyond several hundred MeV.

With the obvious exceptions of $p\bar{p}$ and $e^{\pm}$, massive particles are stable only on non-astrophysical timescales, which are nonetheless significantly longer than the time required to travel to the region where they would interact with the incident radiation. For instance, a neutron emitted from a 100 GeV PBH, has a lifetime of approximately 1 day. MacGibbon [2] has shown that when the decay of primary particles is considered, eq. (16) becomes:

$$\frac{d\dot{N}_X}{dE_{\mathrm{TOT}}} = \sum_j \int_{E_{\mathrm{TOT}}}^{\infty} \frac{\Gamma_j(E,T)}{h}\left[\exp\left(\frac{E}{k_B T_H}\right) - (-1)^{2s_j}\right]^{-1} \cdot \frac{dg_{jX}(E, E_{TOT})}{dE_{TOT}} dE_{TOT}\,, \tag{17}$$

where $\frac{dg_{jX}(E, E_{TOT})}{dE_{TOT}} dE_{TOT}$ is the relative number of particles of species *X* possessing energy $E_{TOT}$ that are created by particle *j* with energy *E*, and $\frac{d\dot{N}_X}{dE_{\mathrm{TOT}}}$ is the instantaneous flux of particles in species *X*. The purpose of the sum of *j* is to account for all energy-carrying particle species and their concomitant degrees of freedom.

The de Broglie wavelength $\lambda_B$ of an interacting particle (hence its effective "size") emitted with an energy equal to the Hawking temperature, is $8\pi^2 R_S$, where $R_S$ is the Schwarzschild radius. If the emissions rate is greater than $\frac{c}{\lambda_B}$, interactions between emitted particles, irrespective of their species, would be expected. Consequently, a PBH would be surrounded by a high-density/high-opacity accretion cloud in the region surrounding the event horizon.

However, Oliensis has shown that less than 0.1% of the emitted particles in the lifetime of a PBH are interacting [8]. Therefore, a PBH's emitted particles are not self-interacting as a result of short-range forces prior to fragmentation. Furthermore, color is irrelevant to the short-range propagation of emitted particles. Thus, a dense cloud of emitted particles does not surround the primordial black hole.

Additionally, relativistic jets of emitted particles are fragmented as a result of $q\bar{q}$ pairs produced in the region $r \sim 1$ of the PBH, with one quark tunneling back to the $r < 1$ region. The color field lines connecting the quark and anti-quark, located in the $r < 1$ and $r > 1$ regions, are compressed into a "conduit-like" length of spacetime. If there is constant linear energy density in the interior of the conduit, the potential energy between the quarks is proportional to their spatial separation. As their separation increases, the potential energy also increases to the value required to produce another $q\bar{q}$ pair, and the color conduit is hewn into two approximately equal length color conduits. This process will continue until the quark and gluon kinetic energies drop below the fragmentation threshold, at which point, color coupling will dominate. This signifies the end of fragmentation and the commencement of hadronization (the grouping of particles into color-singlet states) [9].

If an intense field of relativistic protons is incident upon the horizon of a PBH, multiple interactions with Hawking radiation will occur. Examined here are $p + \nu\bar{\nu}$, $p + p\bar{p}$, $p + e^{\pm}$, and $p + \gamma$ collisions, and their likely effect on the net flux of subsequent incident fermions. While an incident radiation field of any particle species could be considered, protons are selected, and their interactions with all of the emergent Hawking species are described.

Although neutrinos are often treated as non-interacting particles, neutrino-annihilation, neutrino-absorption, and neutrino-nucleon scattering can, in varying degrees, all be confidently expected at the energy levels considered here. Although the formation of an accretion cloud of emitted Hawking particles is not expected, an accretion veil, resulting from the interaction with incident fermions, is anticipated.

### 3.4.1 $\nu\bar{\nu}$ Absorption

Nucleonic absorption of neutrinos and anti-neutrinos results in momentum and energy transfers to the nucleons. Kneller et al. have determined the neutrino-antiproton and neutrino-neutron absorption cross sections to be [10]:

$$\sigma_{\nu n}^{\text{abs}} = \sigma_o \left(\frac{1+3g_A^2}{4}\right)\left(\frac{E_\nu + \Delta}{m_e^2}\right)^2 \sqrt{1-\left(\frac{m_e}{E_\nu + \Delta}\right)^2} W_M \,, \tag{18}$$

and

$$\sigma_{\bar{\nu} p}^{\text{abs}} = \sigma_o \left(\frac{1+3g_A^2}{4}\right)\left(\frac{E_\nu - \Delta}{m_e^2}\right)^2 \sqrt{1-\left(\frac{m_e}{E_{\bar{\nu}} - \Delta}\right)^2} W_{\bar{M}} \,, \tag{19}$$

where $g_A$ and $\Delta$ are the axial-vector coupling constant and the neutron-proton mass difference respectively. $W_M$ and $W_{\bar{M}}$ are the two weak magnetism corrections [11] given by:

$$\text{a} W_M = 1 + 1.1 \frac{E_\nu}{m_n}, \tag{20}$$

$$W_{\bar{M}} = 1 - 7.1 \frac{E_{\bar{\nu}}}{m_n}. \tag{21}$$

The rates of total momentum and energy transfer from neutrinos materializing at $(\theta, \phi)$, $F^{(\text{abs})}$ and $W^{(\text{abs})}$ respectively, per particle, to all nucleonic particles at position $(r, z)$ are:

$$F^{(\text{abs})} = \int d\Omega_\nu \left[-\sin\theta\cos\phi\,\hat{i} + \cos\theta\,\hat{k}\right] \int dE_\nu \frac{b(r_\nu, E_\nu)}{4\pi} S_\nu(\theta_\nu, \phi_\nu) E_\nu \sigma_{\nu N}^{(\text{abs})} \tag{22}$$

$$W^{(\text{abs})} = \int d\Omega_\nu \int dE_\nu \frac{b(r_\nu, E_\nu)}{4\pi} S_\nu(\theta_\nu, \phi_\nu) E_\nu \sigma_{\nu N}^{(\text{abs})}, \tag{23}$$

where $b(r_\nu, E_\nu)$ is the differential neutrino flux per unit area that derives from radial coordinate $r_\nu$ in the frame of the accretion disk. The momentum transfer occurs in the $\hat{k}'$ direction. The basis vectors transformations are:

$$\hat{i}' = -\cos\theta\cos\phi\,\hat{i} - \sin\theta\cos\theta\,\hat{j} - \sin\theta\,\hat{k}, \tag{24}$$

$$\hat{j}' = \sin\phi\,\hat{i} - \cos\phi\,\hat{j}, \text{ and} \tag{25}$$

$$\hat{k}' = -\sin\theta\cos\phi\,\hat{i} - \sin\theta\sin\phi\,\hat{j} + \cos\theta\,\hat{k}. \tag{26}$$

### 3.4.2 $\nu\bar{\nu}$ Annihilation

The energy density deposition rate $\frac{dL_{\nu\bar{\nu}}}{dV}$, for the annihilation of neutrino-antineutrino pairs into electron-positron pairs, is independent of the polar angle,

$$\frac{dL_{\nu\bar{\nu}}}{dV} = \iint d\Omega_\nu d\Omega_{\bar{\nu}} \iint dE_\nu dE_{\bar{\nu}} \frac{b(r_\nu, E_\nu) b(r_{\bar{\nu}}, E_{\bar{\nu}})}{16\pi^2} S_\nu(\theta_\nu, \phi_\nu) S_{\bar{\nu}}(\theta_{\bar{\nu}}, \phi_{\bar{\nu}}) \times \frac{E_\nu + E_{\bar{\nu}}}{E_\nu E_{\bar{\nu}}} \left\{ E_\nu E_{\bar{\nu}} |\mathrm{v}_\nu - \mathrm{v}_{\bar{\nu}}| \sigma_{\nu\bar{\nu}} \right\} \tag{27}$$

where $b(r_{\bar{\nu}}, E_{\bar{\nu}})$ is the differential neutrino flux per unit area that emerges from radial coordinate $r_{\bar{\nu}}$ in the frame of the accretion disk. $\left\{ E_\nu E_{\bar{\nu}} |\mathrm{v}_\nu - \mathrm{v}_{\bar{\nu}}| \sigma_{\nu\bar{\nu}} \right\}$ is a Lorentz invariant, which is most easily determined in the center of mass frame of the accretion disk, and then articulated in terms of *s*, the Mandelstam variable.

Herrera et al. have calculated the annihilation cross section, which allows eq. (27) to be evaluated [12],

$$E_\nu E_{\bar{\nu}} |\mathrm{v}_\nu - \mathrm{v}_{\bar{\nu}}| \sigma_{\nu\bar{\nu}} = \frac{\sigma_o s^2}{48 m_e^2} \sqrt{1 - \frac{4m_e^2}{s}} \left[ 2\left(C_V^2 + C_A^2\right) + \frac{4m_e^2}{s}\left(C_V^2 - 2C_A^2\right) \right], \tag{28}$$

where $C_V = \frac{1}{2} + 2\sin^2\theta_W$, $C_A = \frac{1}{2}$, $\sigma_o = \frac{4G_F^2 m_e^2}{\pi\hbar^4}$, and $\theta_W = \cos^{-1}\frac{m_W}{m_Z}$ is the Weinberg Angle.

$S_\nu(\theta_\nu, \phi_\nu) = \Theta(a_\nu - r_\nu) + \Theta(r_\nu - a_\nu)\sec\theta_\nu$ and $S_{\bar{\nu}}(\theta_{\bar{\nu}}, \phi_{\bar{\nu}}) = \Theta(a_{\bar{\nu}} - r_{\bar{\nu}}) + \Theta(r_{\bar{\nu}} - a_{\bar{\nu}})\sec\theta_{\bar{\nu}}$ are the corrections to the neutrino and antineutrino differential number fluxes per unit area respectively, and which account for neutrino and antineutrino trapping occurring within the optically thick and optically thin zones of the disk. $\Theta$ is the Heaviside function. $a_\nu$ and $a_{\bar{\nu}}$ are the neutrino and antineutrino radial boundaries respectively.

For stellar mass black holes, $a_\nu$ and $a_{\bar{\nu}}$ have values from several 10s to several 100s of km [10]. Due to the significantly smaller radial boundaries in a PBH, neutrino and antineutrino trapping within the accretion disk is expected to be greatly reduced, and the contributions of $S_\nu(\theta_\nu, \phi_\nu)$ and $S_{\bar{\nu}}(\theta_{\bar{\nu}}, \phi_{\bar{\nu}})$ are likely inconsequential.

#### 3.4.3 $\nu\bar{\nu}$ Scattering

Momentum and energy transfer, due to the scattering of neutrinos and antineutrinos by protons within the incident fermion field, are given by:

$$F = \int d\Omega_\nu \left[ -\sin\theta\cos\phi\,\hat{i} + \cos\theta\,\hat{k} \right] \int dE_\nu \frac{b(r_\nu, E_\nu)}{4\pi} S_\nu(\theta_\nu, \phi_\nu) \int d\Omega'_\nu\, p_{k'} \frac{d\sigma}{d\Omega'_\nu}, \tag{29}$$

$$W = \int d\Omega_\nu \int dE_\nu \frac{b(r_\nu, E_\nu)}{4\pi} S_\nu(\theta_\nu, \phi_\nu) \int d\Omega'_\nu\, T \frac{d\sigma}{d\Omega'_\nu}, \tag{30}$$

where $p_{k'} = \frac{E_\nu(E_\nu + M)(1 - \cos\theta_{\nu'})}{M + E_\nu(1 - \cos\theta_{\nu'})}$, and $T = \frac{E_\nu^2(1 - \cos\theta_{\nu'})}{M + E_\nu(1 - \cos\theta_{\nu'})}$.

The mass of the scattered particle, in this case a proton, is *M*. The neutrino is scattered at angle $\theta_{\nu'}$, measured with respect to basis vector $\hat{k}'$. The momentum transfer is always independent of the mass of the scattered particle, and is approximately equal to the neutrino energy.

In the event that $M >> E_\nu$, the energy transfer is negligible. However, as Table 4 shows, the energy of emitted neutrinos and antineutrinos for a 100 GeV PBH is 2.829 GeV (~3 times the proton rest mass). Consequently, the energy transfer from $\nu\bar{\nu}$ scattering is not insignificant.

The differential cross section for neutrino-proton scattering is [13]:

$$\frac{d\sigma_{\nu p}}{d\Omega_\nu} = \frac{\sigma_o}{16\pi}\left(\frac{E_\nu}{m_e}\right)^2\left[(C_V - 1)^2 + 3g_A^2(C_A - 1)^2\right]\left(1 + \delta_p \cos\theta'_\nu\right), \qquad (31)$$

where $\delta_p = \frac{(C_V - 1)^2 - g_A^2(C_A - 1)^2}{(C_V - 1)^2 + 3g_A^2(C_A - 1)^2}$.

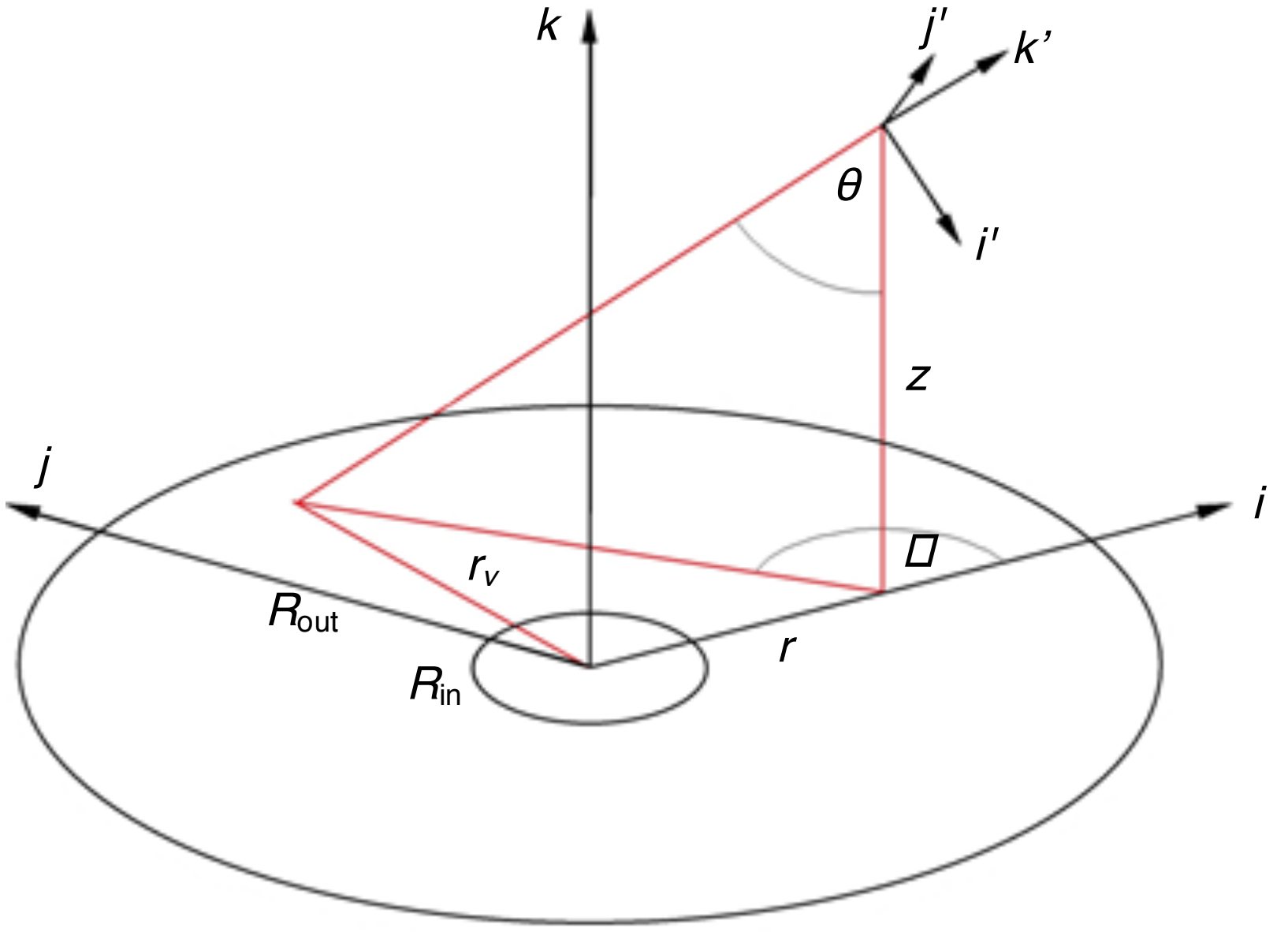

Figure 3: Basis vectors, distances and angles for neutrino absorption and scattering. The singularity is at the origin. $R_{out}$ and $R_{in}$ refer to the outer and inner accretion disk radii respectively. $i'$ is within the same plane as $\theta$. [10]

### 3.4.4 Critical Density

It is clear that the energy deposition resulting from neutrino-antineutrino annihilation into electron-positron pairs, as well as the scattering of neutrinos, will impede the flux of nucleons inbound to the PBH. Fryer & Mészáros [14] suggest the following approach to the problem of a differential mass element, beginning at infinity, approaching a stellar mass black hole. This approach is also valid for a fermion field incident upon a PBH. A region of the field with density $\rho$ will experience a gravitational force $F_g$. It will further experience a force resulting from neutrino annihilation $F_{\nu\bar{\nu}}$ and from both scattering and absorption $F_{\nu}$. Since the fermion field emerges from infinity (i.e., $r >> 1$), the change in a mass element's kinetic energy between infinity and a height *z* is:

$$\frac{1}{2}\rho v^2(z) = \int_z^{\infty} dz \left(F_g + F_{\nu} + F_{\nu\bar{\nu}}\right). \tag{32}$$

The gravitational force is:

$$F_G = \frac{-G_N M_{\mathrm{SK}} \rho}{z^2} \tag{33}$$

The force responsible for the neutrino-antineutrino annihilation acceleration is:

$$F_{\nu\bar{\nu}} = \frac{1}{c}\frac{dL_{\nu\bar{\nu}}}{dV} \tag{34}$$

The force responsible for the acceleration due to both neutrino-antineutrino scattering and absorption is:

$$F_{\nu} = n_n\left(F_n^{(\mathrm{abs})} + F_{n\nu} + F_{n\bar{\nu}}\right) + n_p\left(F_p^{(\mathrm{abs})} + F_{p\nu} + F_{p\bar{\nu}}\right) + n_e\left(F_{e\nu} + F_{e\bar{\nu}}\right), \tag{35}$$

where $n_i$ are the number densities. In the case of atomic matter falling into a black hole, $n_p = n_e$ and $\rho = (n_n + n_p)m_u$; $m_u$ is the atomic mass unit.

Fryer & Mészáros introduce a neutron fraction, *Y*, given by:

$$Y = \frac{n_n}{n_n + n_p}. \tag{36}$$

Consequently, eq. (35) becomes:

$$F_{\nu} = \frac{\rho}{m_u}\left[Y F_n + (1 - Y)\left(F_p + F_e\right)\right]. \tag{37}$$

$F_n$, $F_p$ and $F_e$ are the total momentum transfer rates. Since $F_G$ and $F_{\nu}$ are functionally dependent on $\rho$, there must exist a value of $\rho = \rho_o$ at which:

$$\int_z^{\infty} dz \left(a_g + a_{\nu} + a_{\nu\bar{\nu}}\right) = 0, \tag{38}$$

$$\rho_o = \frac{\int_z^\infty dz \frac{1}{c} \frac{dL_{\nu\bar{\nu}}}{dV}}{\int_z^\infty dz \left[ \frac{G_N M_{SK}}{z^2} - \frac{YF_n + (1-Y)(F_p + F_e)}{m_u} \right]}. \tag{39}$$

If $\rho < \rho_o$, an incoming mass element will be ejected at a height greater than *z*. If $\rho = \rho_o$, an incoming mass element will be ejected at *z*. However, if $\rho > \rho_o$, the mass element will continue past *z*. Kneller et al. have named the maximum value of $\rho_o$, the critical density $\rho_*$. Material with a density greater than the critical density cannot be neutrino- or antineutrino-ejected at any value of *z*, and is therefore accreted into the black hole.

In the case of a proton radiation field, incident upon a PBH, $F_n = F_e = 0$, $F_p = 1$, and $Y = 0$. Eqs. (37) and (39) reduce to:

$$F_\nu = \frac{\rho}{m_u} \tag{40}$$

and

$$\rho_o = \frac{\int_z^\infty dz \frac{1}{c} \frac{dL_{\nu\bar{\nu}}}{dV}}{\int_z^\infty dz \left[ \frac{G_N M_{\mathrm{PBH}}}{z^2} - \frac{1}{m_u} \right]}. \tag{41}$$

Using the Schwarzschild radius and infinity as limits of integration guarantees that $p\nu$ and $p\bar{\nu}$ interactions will not prevent protons from being accreted into the PBH. Thus:

$$\rho_o = \frac{\int_{R_S}^\infty dz \frac{1}{c} \frac{dL_{\nu\bar{\nu}}}{dV}}{\int_{R_S}^\infty dz \left[ \frac{G_N M_{\mathrm{PBH}}}{z^2} - \frac{1}{m_u} \right]}. \tag{42}$$

### 3.4.5 $pp$ and $p\bar{p}$ Scattering

An intense fermion radiation field will strongly resemble a highly collimated, incident beam of nearly infinite pseudorapidity protons, and will undergo hard scattering with the protons emerging as Hawking radiation. Since the effective radius of all of these protons is less than the Schwarzschild radius (and much less than the classical proton radius), the interaction is best represented as collisions between the constituent partons. The cross section calculation consists of terms containing the partonic scatter cross section $\hat{\sigma}$, and the parton density functions $f_{i,p}$, and is given by [15]:

$$\sigma(pp \to X) = \sum_{i,j} \int dx_1 dx_2 f_{i,p}(x_1, \mu_F^2) f_{i,p}(x_2, \mu_F^2) \hat{\sigma}_{ij \to X}(x_1 x_2 s, \mu_R^2, \mu_F^2) \quad (43)$$

The sum over *i* and *j* is to account for all initial-state partons with longitudinal momentum fractions $x_1$ and $x_2$, capable of giving rise to the final state *X* whose center of mass energy is $\sqrt{x_1 x_2 s}$. $\mu_R^2$ and $\mu_F^2$ are the factorization scales, which are recovered from truncations of the expansion of the strong coupling constant, and which yield universal parton densities at a given resolution.

Parton-parton scattering is arguably by far, the most frequent hadron collision process that would occur between a fermion radiation field and Hawking radiation protons. Shortly after being collisionally created (and in many cases, prior to reaching the horizon), hard partons would continuously radiate low-energy collinear gluons as a *parton shower*. If a high energy scattering of two protons occurs at a significant distance from the event horizon of the PBH, the emitted high-energy parton will reach distances from the proton constituents which are much larger than the constituents' effective radii; as a result, an increase in the QCD force would occur. Radiation of continuously softer gluons, at small angles relative to the initial parton, will likely continue.

When this occurs in a weak gravitational field (i.e., at $r > 1$), eventually, a non-perturbative transition would form color-neutral hadrons as a result of parton binding. A reasonably well-collimated hadron jet would ensue; its total energy and momentum would be comparable to the original scattered parton. If this process remains applicable in the vicinity of the horizon (i.e., $r \to 1$), then relatively little energy loss by incoming protons would be expected. Most of the incident energy would be transported by the post-hadronization hadronic jet. The acollinearity effect resulting from the emission of soft gluons reduces the probability of further collisions. However, if the extreme gravitation gradient in the $r \to 1$ region prevents parton binding, and consequently color-neutral hadrons do not form, an uncollimated parton shower could be accreted in the spacetime region surrounding the horizon.

The Compact Muon Solenoid experiment at the LHC produced hadron collisions with a double-differential inclusive jet cross section between approximately $10^{-2}$ and $10^{7}$ pb/GeV for $200 \leq p_T(\text{GeV}) \leq 1000$, $L = 34\text{pb}^{-1}$, $\sqrt{s} = 7\text{TeV}$, and $2.5 \leq |y| < 15625$ [15].

The scattering picture for proton-antiproton collisions is exceptionally similar. The collision cross section is:

$$\sigma(p\bar{p} \to X') = \sum_{i,j} \int dx_1 dx_2 f_{i,p}(x_1, \mu_F^2) f_{i,\bar{p}}(x_2, \mu_F^2) \hat{\sigma}_{ij \to X'}(x_1 x_2 s, \mu_R^2, \mu_F^2). \quad (44)$$

In low-gravity environments, the cross sections for proton-antiproton collisions with $\sqrt{s} \gtrsim 100$ GeV are approximately equal [16]. Low energy ($\sqrt{s} \lessgtr 10$ GeV) proton-proton collisions have cross sections which are 2-3 times smaller than equivalent-energy proton-antiproton collision cross sections.

### 3.4.6 $pe^{\pm}$ Scattering

The above-discussed proton field will experience deep inelastic scattering with Hawking radiation electrons and positrons. This results in electron-quark fusion creating resonance peaks in the electron-proton collision cross sections, and potentially, hypothetical scalar leptoquark (LQ) isodoublet production. Considered here is the reaction:

$$e^{\pm} + q \rightarrow \gamma + LQ\,, \tag{45}$$

which comes from:

$$e^{\pm} + p \rightarrow \gamma + LQ + X \tag{46}$$

in which the LQ interacts only with the first generation fermions.

The integrated cross section of the electron-proton collision is determined by convoluting the differential cross sections of the hard subprocesses with the related parton density functions [17]:

$$\sigma(s) = \int_{x_{\min}}^{1} dx q(x, Q^2) \int_{-1}^{1} d\cos\vartheta_{\gamma} \cdot \frac{d\hat{\sigma}(\hat{s}, \cos\vartheta_{\gamma})}{d\cos\vartheta_{\gamma}} \cdot \Theta_{\text{cuts}}(E_{\gamma}, \vartheta_{\gamma}), \tag{47}$$

where $q$ is a constituent quark of a targeting proton, $\hat{\sigma}$ is the cross section of eq. (46), $q(x, Q^2)$ is the quark distribution function, the 4-momentum transfer scale is $Q^2 = \hat{s}$, $\hat{s} = xs$, $\vartheta_{\gamma}$ is the photon emission angle relative to the proton beam, and $\Theta_{\text{cuts}}(E_{\gamma}, \vartheta_{\gamma})$ accounts for the necessary kinematical cuts. Since eq. (45) is infrared divergent, $E_{\gamma} > E_{\gamma}^{0} > 0$.

A Monte Carlo simulation with $E_{\gamma}^{0} = 1$ GeV (equivalent to the $e^{\pm}$ energy radiated from a 10 GeV PBH), a photon emission angle cut of $\vartheta_{\gamma}^{\min} < \vartheta_{\gamma} < \vartheta_{\gamma}^{\max}$, a center of mass energy of 1740 GeV (corresponding to a 7.6 TeV proton beam), an electron energy of 100 GeV (much greater than the $e^{\pm}$ energy radiated from any considered PBH), and an integrated luminosity of 1 $\text{fb}^{-1}$ [17] does not imply that the formation of leptoquarks would dissipate significant energy of incident protons. Even less proton field energy should be dispelled toward leptoquark formation when incident upon a 100 GeV PBH, since the center of mass energy for a ~7.6 TeV proton field would be ~320 GeV.

### 3.4.7 $p\gamma$ Scattering

$p\gamma$ scattering in the vicinity of a PBH is likely to be diffractive and in the form of $\gamma p \rightarrow \gamma X$. Perturbative calculations of the scattering cross section at large $t$ and extreme energies $(W^2 >> |t| >> \Lambda_{\text{QCD}}^2)$ exceed the $\frac{J}{\Psi}$ photoproduction cross section, and are therefore considered here. The complete scattering cross section is [18]:

$$\frac{d\sigma(\gamma q \rightarrow \gamma q)}{dt} = \frac{|A_{(+,+)}|^2 + |A_{(+,-)}|^2}{16\pi s^2}, \tag{48}$$

where the amplitudes $A_{(+,+)}$ and $A_{(+,-)}$ are:

$$A_{(+,+)} = i\frac{6}{9}\alpha_{em}\alpha_s^2\frac{4\pi}{3}\frac{s}{|t|}\int d\nu \frac{\nu^2}{\left(\frac{1}{4}+\nu^2\right)^2}\left(\frac{\frac{11}{4}+3\nu^2}{1+\nu^2}\right)\left(\frac{\tanh(\pi\nu)}{\pi\nu}\right)\left(\frac{s}{|t|}\right)^{\omega(\nu)}, \quad (49)$$

$$A_{(+,-)} = i\frac{6}{9}\alpha_{em}\alpha_s^2\frac{4\pi}{3}\frac{s}{|t|}\int d\nu \frac{\nu^2}{\left(\frac{1}{4}+\nu^2\right)^2}\left(\frac{\frac{1}{4}+\nu^2}{1+\nu^2}\right)\left(\frac{\tanh(\pi\nu)}{\pi\nu}\right)\left(\frac{s}{|t|}\right)^{\omega(\nu)}, \quad (50)$$

for

$$\omega(\nu) = \frac{3\alpha_s}{\pi}\left[2\Psi(1) - \Psi\left(\frac{1}{2}+i\nu\right) - \Psi\left(\frac{1}{2}-i\nu\right)\right]. \quad (51)$$

Saddle point approximations for eqs. (49) and (50) yield:

$$A_{(+,+)} = \frac{528}{27}i\alpha_{em}\alpha_s^2\frac{s}{|t|}\left(\frac{\pi}{7\varsigma(3)\eta}\right)4^{\eta}, \quad (52)$$

and

$$A_{(+,-)} = \frac{48}{27}i\alpha_{em}\alpha_s^2\frac{s}{|t|}\left(\frac{\pi}{7\varsigma(3)\eta}\right)^{\frac{3}{2}}4^{\eta} \quad (53)$$

where:

$$\eta \equiv \frac{6}{\pi}\alpha_s \ln\left(\frac{s}{|t|}\right). \quad (54)$$

$\varsigma(\nu)$ is the Riemann zeta function. $t$ is the squared momentum transfer. For a more detailed explanation, consult [18].

A Vector Dominance Model numerical simulation performed by Ivanov and Wusthoff [18] showed:

$$\frac{d\sigma(\gamma)}{dt} \approx 7.5nb\exp\left(-5.3|t|\,\mathrm{GeV}^{-2}\right). \quad (55)$$

For the case of Hawking photons emitted from a 100 GeV PBH scattering off an incident 2 TeV proton field, extrapolation of the Ivanov and Wusthoff simulation data yields a cross section of approximately 8.2 pb. If a 7.6 TeV proton field (as considered in the $p\gamma$ Scattering section) is incident, the cross section is approximately 3.6 pb.

### 3.4.8 Interaction Summary

The above analyses of interaction processes and scattering cross sections of incident fermions and PBH Hawking radiation have phenomenologically shown scattering cross sections that are significantly smaller than the available Schwarzschild targeting cross section $\left(4\pi R_S^2\right)$. For TeV proton fields incident upon a 100 GeV PBH, the Schwarzschild targeting cross section is four orders of magnitude larger than the proton-gamma-ray interaction cross section.

However, Hawking particle fluxes range from $10^{26}$ – $10^{28}$ GeV$^{-1}$·s$^{-1}$ for a 100 GeV PBH (Table 3). If equivalent incident fluxes are isotropically distributed across $2\pi$ sr of the horizon, the interaction cross section can increase volumetrically by as much as 26-28 orders of magnitude, yielding a "volumetric" cross section of potentially 100 Tb - 10 Pb. In this event, a [relatively] enormous accretion cloud would surround the PBH. The approximately equivalent stellar scale would be an accretion disk 30,000 AU - 300,000 AU in radius surrounding a $10M_\odot$ black hole[ii].

## 4. Extrapolation to Planck-scale PBHs

While the flux of Hawking radiation from Planck-scale primordial black holes is uncertain, the spectrum can be taken to consist of particles with the Planck energy $\sqrt{\frac{\hbar c^5}{8\pi G_N}} \sim 2.43\times 10^{18}\ \mathrm{eV}$.
At the Planck time prior to evaporation, the PBH temperature is the Planck temperature $T_P = \sqrt{\frac{\hbar c^5}{G_N k_B^2}} \sim 1.42\times 10^{32}\ \mathrm{K}$. In this context, eq. (15) becomes:

$$P_{\mathrm{abs}} = 1 - \exp\left(-\frac{\varepsilon}{T_P}\right) \tag{56}$$

Since no incident particles can have energy in excess of the Planck energy, the maximum absorption probability from eq. (56), discounting the interaction with Hawking radiation, is $1-\exp(-1) \sim 0.63$.

While the final evaporation state of a primordial black hole remains unresolved, the expected time required for a Planck-sized black hole to evaporate is the Planck time. Therefore, only photons reaching the horizon at the Planck time would have a non-zero probability of being absorbed before total evaporation. However, since all particles reaching the horizon would first have to traverse the emergent Hawking radiation and the intense parton showers of previous collisions, energy dissipation from scattering and annihilation of all incoming particles would be expected. Thus, all incoming radiation, even particles with the Planck energy, will experience a reduction in energy by means of collisions and scattering with both the expelled Hawking radiation and the resulting quark- and gluon-rich accretion cloud. Therefore, it is expected that no particles could arrive in the vicinity of the PBH horizon with the Planck energy.

---

[ii] These radii are the distances from the sun to the Spherical Oort Cloud and Outer Oort Cloud respectively.

However, since the absorption probability is asymptotic toward zero, absorption of incident particles by the PBH cannot be entirely ruled out. Thus, even though relatively low energy particles have an exceedingly negligible chance of being absorbed by a Planck-sized PBH, the absorption probability is not actually zero.

Consequently, for their Planck time lifetimes, Planck-sized primordial black holes best approximate white holes, in so far as their near 1 scattering probability is concerned. However, a totally opaque collisionally-produced accretion cloud will not occur, and a state of “absolute whiteness” of primordial black holes is not possible.

## Summary

The Hawking radiation spectrum of attometer primordial black holes has been described, and the non-interactive absorption probability of incident fermions was discussed. Although the non-interactive absorption probability of incident fermions by a PBH is not insignificant, the probability of a fermion incident on the horizon being absorbed is substantially smaller when the interaction with Hawking radiation is considered.

When the Schwarzschild radius of a primordial black hole is extrapolated to the Planck length, even incident fermions with the Planck energy have only an extremely negligible chance of absorption due to the high opacity of the surrounding accretion cloud and the need to cross the intervening distance in the Planck time. Subsequently, as primordial black holes evaporate and their Schwarzschild radii approach the Planck length, they asymptotically mimic white holes. However, the scattering-probability-equal-to-one characteristic of a white hole is never actually achieved by a PBH.

A more confident picture of $pp$, $p\bar{p}$, $pe^{\pm}$, $p\gamma$, and $p\nu\bar{\nu}$ interactions involving quark-gluon scattering in the purlieu of the horizon of a Planck-sized or near Planck-sized PBH awaits further developments in Quantum Gravity.

# References

[1] L. Crane and S. Westmoreland, "ARE BLACK HOLE STARSHIPS POSSIBLE?," *arXiv:0908.1803 [gr-qc] ,* 12 August 2009 Last Accessed: 13th October 2013.

[2] J. MacGibbon and B. Webber, "Quark- and gluon-jet emission from primordial black holes. II. The emission over the black-hole lifetime," *Physical Review D.,* vol. 44, no. 2, pp. 376-392, 1991.

[3] V. P. Frolov and I. D. Novikov, Black Hole Physics: Basic Concepts and New Developments, Kluwer Academic Publishers, 1997, p. 398.

[4] R. Sini and V. Kuriakose, "Absorption cross section and Emission spectra of Schwarzschild black hole in Dirac field.," *Mod. Phys. Lett.,* vol. A, no. 23, pp. 2867-2879, 2008.

[5] S. Hawking, *Sci. Am.,* vol. 236, no. 34, 1977.

[6] N. Birrell and P. Davies, "Quantum Fields in Curved Space," *Cambridge University Press,* p. 263, 1982.

[7] M. Perry, *Phys. Lett.,* vol. 71B, no. 234, 1977.

[8] J. Oliensis and T. Hill, *Phys. Lett.,* vol. 143B, no. 92, 1984.

[9] F. Halzen and A. Martin, Quarks and Leptons: An Introductory Course in Modern Particle Physics, New York: Wiley, 1984.

[10] J. Kneller, G. McLaughlin and R. Surman, "Neutrino Scattering, Absorption and Annihilation above the accretion disks of Gamma Ray Bursts," *arXiv:astro-ph/0410397v1 ,* 2008 Last Accessed: 13th October 2013.

[11] C. Horowitz, "Weak magnetism for antineutrinos in supernovae," *arXiv:astro-ph/0109209,* vol. 65, 9 October 2001 Last Accessed: 13th October 2013.

[12] M. Herrera and S. Hacyan, *ApJ,* vol. 336, no. 539, 1989.

[13] A. Burrows, S. Reddy and T. A. Thompson, "NEUTRINO OPACITIES IN NUCLEAR MATTER," *arXiv:astro-ph/0404432,* 2004 Last Accessed: 13th October 2013.

[14] C. Fryer and P. Mészáros, *ApJ,* vol. 588, no. L25, 2003.

[15] J. M. Butterworth, G. Dissertori and G. P. Salam, "Hard Processes in Proton-Proton Collisions at the Large Hadron Collider," *arXiv:1202.0583 ,* 2 February 2012 Last Accessed: 13th October 2013.

[16] R. Godbole, A. Grau, G. Pancheri and Y. Srivastava, "Hunting for asymptotia at LHC," *Phys. Rev. D.,* vol. 72, no. 076001, 2005.

[17] V. Ilyin, A. Pukhov and V. Savrin, "Single leptoquark production associated with hard photon emission in ep collisions at high energies," *arXiv:hep-ph/9503401v2 Last Accessed: 13th October 2013,* 17 May 1995.

[18] D. Y. Ivanov and M. Wusthoff, "Hard diffractive Photon-Proton Scattering at large t," *Eur. Phys. J,* vol. C8, pp. 107-114, 1999.

[19] C. Amsler, "(Particle Data Group)," *Physics Letters,* vol. B667, 2008.

[20] N. R. E. Laboratory, *Reported timeline of solar cell energy conversion efficiencies,* 2007.

[21] M. Gruntman, "Imaging the three-dimensional solar wind," *Journal of Geophysical Research,* vol. 106, no. A5, pp. 8205-8216, 1 May 2001.

[22] V. Vasyliunas and G. Siscoe, "On the Flux and the Energy Spectrum of Interstellar Ions in the Solar System," *Journal of Geophysical Research,* vol. 81, no. 7, pp. 1247-1252, 1 March 1976.

[23] Y. Gu, Z. Zhu, Y. Li, C. Chen, Q. Kong and S. Kawata, "Collimated GeV proton beam generated by the interaction of ultra-intense laser with a uniform near-critical underdense plasma," *EPL,* vol. 95, August 2011.

[24] A. Maksimchuk, S. Gu, K. Flippo, D. Umstadter and V. Bychenkov, "Forward ion acceleration in thin films driven by a high-intensity laser," *Phys. Rev. Lett.,* vol. 84, no. 4108, 2000.

[25] O. Buneman, "Dissipation of currents in ionized media," *Phys. Rev.,* vol. 115, pp. 503-517, 1959.

[26] X. Yan, H. Wu, Z. Sheng, J. Chen and J. Meyer-ter-Vehn, "Self-organizing GeV nano-Coulomb collimated proton beam from laser foil interaction at 7 × 1021 W/cm2," *Phys. Rev. Lett.,* vol. 100, 2008.

[27] S. Hawking, A Brief History of Time, New York: Bantam Books, 1988.